\documentstyle[aps,prl,multicol,psfig]{revtex}
\tighten
\begin{document}

\begin{multicols}{2}

%
%

{\bf Comment on ``Insulator-to-Metal Crossover in the Normal State
of $\rm La_{2-x}Sr_xCuO_4$ Near Optimum Doping''}

%

In a recent letter Boebinger {\it et al.}  \cite{boebinger} 
report the results of transport experiments
in ${\rm La_{2-x}Sr_xCuO_4}$ (LSCO), in the presence of high
magnetic fields, allowing to access the normal phase
underlying the superconducting region. They assess the
presence of a metal-insulator transition in the underdoped 
region ending near optimal doping ($x=0.17$) at $T=0$.
In this comment, we point out that their work not only
shows the existance of 
a quantum critical point (QCP) {\it different from the} 
antiferromagnetic (AF) QCP, but also 
provides a clear evidence that the  nature of this
instability is related to charge-ordering. We also
argue that a careful analysis of their data allows to locate the
low-temperature metal-insulator transition, and therefore the
QCP, at larger doping ($x\approx 0.2$) in closer agreement with
the zero-temperature extrapolation of the 
pseudogap data reported in Ref. \cite{batlogg}.

Several experimental findings give indirect support 
for the existence of a QCP at (or near) optimal doping:
neutron scattering, optical spectroscopy, NMR, 
susceptibility, photoemission, specific heat, thermoelectric power,
Hall coefficient, resistivity  
(see, e.g., the discussion in Ref. \cite{evora}).
The resistivity measurements in Ref. \cite{boebinger}
are the first direct evidence of a
transition ending near optimal doping at $T=0$.
On the other hand, the origine of this transition to an insulating
state is less easily established.
A crucial hint, not considered by the authors,
 is provided by the observation that the
resistivity curve for $x=0.12$, i.e. near the "magic" doping 
$1/8$, displays an insulating behavior at a much higher temperature
than for values of $x$ immediately nearby.
This shows that commensurability plays a relevant role
in establishing the insulating phase, thereby indicating that
spatial order of the charge degrees of freedom should be involved
also away from commensurability.
This would also agree with the observation made in Ref. 
\onlinecite{boebinger} that the system is rather clean
($k_F {l} \sim 13$)
and the disorder cannot be the source of the transition.
 In this regard a crucial test
would be the sensitivity of the resistivity to non-linear
effects of strong electric fields possibly orienting the 
incommensurate charge-density-wave (ICDW, stripe) domains
thus reducing the boundary mismatches of ordered domains.

Concerning the location of the QCP, we notice that,
on the basis of general arguments of the theory of QCP \cite{evora},
the phase diagram of Fig. 3 in Ref.\onlinecite{boebinger} should be
related to the pseudogap behavior for many physical quantities
pointed out in Ref. \onlinecite{batlogg}.
From this latter analysis the critical point would be located
at a doping value $x\approx 0.2$ larger than the value
($x\approx 0.17$) indicated in Fig.3 of Ref. \onlinecite{boebinger}.
However, this discrepancy can be solved by noting that 
the resistivity curve at $x=0.17$ is likely still affected by 
superconductivity effects, and the metal-insulator transition
at this filling seems to occur at a still finite temperature
(of about 15 K). Therefore, 
the published data of Fig. 1(b) in Ref. \onlinecite{boebinger}
are compatible with a shift of the 
metal-insulator transition at $T=0$, 
towards dopings that are larger than the value $x=0.17$ 
assigned by the authors, in closer
agreement with the analysis of Ref. \onlinecite{batlogg}.

An additional observation can be made by contrasting the behavior of
LSCO systems with La-doped ${\rm Bi_2Sr_2CuO_y}$ (Bi-2201) materials
\cite{ando}. These latter systems are overdoped, making 
quite natural to locate them on the overdoped (quantum disordered)
metallic side of the QCP. Therefore it is not surprising that 
the low temperature behavior of the planar resistivity is always metallic
in Bi-2201. 
On the other hand they are much more anisotropic than the LSCO systems.
Therefore it is again natural to find that $\rho_c$ is always 
semiconducting in  Bi-2201. This strongly twodimensional
character also accounts for the robust linear behavior of the resisitivity
which is expected in the quantum-critical region above 
a twodimensional QCP.
On the contrary, the transverse hopping, being larger 
in LSCO, easily becomes coherent by increasing
doping, giving rise to threedimensional metallic behavior.
In this case the observed $T^{3/2}$ behavior for the planar resistivity
in the overdoped regime 
is explained as the result of the quantum critical 
behavior in three dimension \cite{moriya}.
 
The presence of a QCP ruling
the physical properties of the superconducting cuprates
was repeatedly suggested \cite{pines,varma,CDG,evora}.
 In the proximity of a QCP, critical fluctuations 
can mediate singular interactions between the quasiparticles,
providing both a strong pairing mechanism and a source of 
normal-state anomalies \cite{pines,varma,CDG}. As far as the nature of this
critical point is concerned,
the beautiful experiment by Boebinger and coworkers
appears as a direct experimental confirmation of the
theoretical proposal of the existance of the 
ICDW-QCP \cite{CDG,evora}.

%

Support of the INFM, PRA 1996 is acknowledged.

\vspace {0.2 truecm}

{\small
{C. Castellani, C. Di Castro, and M. Grilli}

Istituto di Fisica della Materia e
Dipartimento di Fisica

Universit\`a di Roma ``La Sapienza''

P.zza A. Moro 2, 00185 Roma, Italy,
 grilli@roma1.infn.it
}

\vspace {-0.5 truecm}
%
%

\end{multicols}
\end{document}